\documentclass[aip,apl,twocolumn,reprint,showpacs,superscriptaddress]{revtex4-1}
\usepackage{amsmath}
\usepackage{amssymb}
\usepackage{graphicx}
\usepackage{hyperref}
\usepackage{url}
\usepackage{color,xcolor}
\begin{document}

\title{Direct observation of ferroelectricity in Ca$_3$Mn$_2$O$_7$ and its prominent light absorption}
\author{Meifeng Liu}
\affiliation{Institute for Advanced Materials, Hubei Normal University, Huangshi 435002, China}
\affiliation{Laboratory of Solid State Microstructures and Innovative Center of Advanced Microstructures, Nanjing University, Nanjing 210093, China}
\author{Yang Zhang}
\author{Ling-Fang Lin}
\affiliation{School of Physics, Southeast University, Nanjing 211189, China}
\author{Lin Lin}
\affiliation{Laboratory of Solid State Microstructures and Innovative Center of Advanced Microstructures, Nanjing University, Nanjing 210093, China}
\author{Shengwei Yang}
\affiliation{Hefei National Laboratory for Physical Sciences at the Microscale, Department of Physics, University of Science and Technology of China, Hefei 230026, China}
\author{Xiang Li}
\affiliation{Institute for Advanced Materials, Hubei Normal University, Huangshi 435002, China}
\affiliation{Laboratory of Solid State Microstructures and Innovative Center of Advanced Microstructures, Nanjing University, Nanjing 210093, China}
\author{Yu Wang}
\affiliation{Institute for Advanced Materials, Hubei Normal University, Huangshi 435002, China}
\author{Shaozhen Li}
\affiliation{School of Physics and Institute for Quantum Materials, Hubei Polytechnic University, Huangshi 435003, China}
\author{Zhibo Yan}
\affiliation{Laboratory of Solid State Microstructures and Innovative Center of Advanced Microstructures, Nanjing University, Nanjing 210093, China}
\author{Xiuzhang Wang}
\affiliation{Institute for Advanced Materials, Hubei Normal University, Huangshi 435002, China}
\author{Xiao-Guang Li}
\affiliation{Hefei National Laboratory for Physical Sciences at the Microscale, Department of Physics, University of Science and Technology of China, Hefei 230026, China}
\author{Shuai Dong}
\email{sdong@seu.edu.cn}
\affiliation{School of Physics, Southeast University, Nanjing 211189, China}
\author{Jun-Ming Liu}
\affiliation{Institute for Advanced Materials, Hubei Normal University, Huangshi 435002, China}
\affiliation{Laboratory of Solid State Microstructures and Innovative Center of Advanced Microstructures, Nanjing University, Nanjing 210093, China}
\affiliation{Institute for Advanced Materials and Laboratory of Quantum Engineering and Materials, South China Normal University, Guangzhou 510006, China}
\date{\today}

\begin{abstract}
Layered perovskites $A_3M_2$O$_7$ are known to exhibit the so-called hybrid improper ferroelectricity. Despite experimentally confirmed cases (e.g. nonmagnetic $M$=Ti and Sn), the ferroelectricity in magnetic Ca$_3$Mn$_2$O$_7$ remains a puzzle. Here, the structural, ferroelectric, magnetoelectric, and optical properties of Ca$_3$Mn$_2$O$_7$ are systematically investigated. Switchable polarization is directly measured, demonstrating its ferroelectricity. In addition, magnetoelectric response is also evidenced, implying the coupling between magnetism and ferroelectricity. Furthermore, strong visible light absorption is observed, which can be understood from its electronic structure. Its direct and appropriate band gap, as well as wide conducting bands, makes Ca$_3$Mn$_2$O$_7$ a potential candidate for ferroelectric photoelectric applications.
\end{abstract}
\maketitle

The so-called $327$-type Ruddlesden-Popper perovskites with generic chemical formula $A_3M_2$O$_7$ ($A$: rare earth or alkaline earth, $M$: transition metal) have been attracting great research attentions since the prediction of hybrid improper ferroelectricity in 2011.\cite{Benedek:Prl} In these hybrid improper ferroelectrics, the ferroelectric polar mode couples with and is driven by other nonpolar modes of structural distortions.\cite{Benedek:Prl,Bousquet:Nat,Rondinelli:Am12,Benedek:Jssc,Mulder:Afm,Benedek:Dt, Pitcher:Sci} Such hybrid improper ferroelectricity not only expands the scope of ferroelectric materials but also provides potential functions like electric-control of magnetization.

In this category, the first predicted two materials are Ca$_3$Ti$_2$O$_7$ and Ca$_3$Mn$_2$O$_7$.\cite{Benedek:Prl} Structurally, these two materials are very similar, as shown in Fig. S1 of supplementary material. The condensation of oxygen octahedral rotation and tilting modes leads to the $A2_1am$ space group as the ground state, whose polarization ($P$) points along the $a$-axis. The predicted $P$'s were $\sim20$ $\mu$C/cm$^2$ for Ca$_3$Ti$_2$O$_7$ and $\sim5$ $\mu$C/cm$^2$ for Ca$_3$Mn$_2$O$_7$.\cite{Benedek:Prl} The intrinsic magnetoelectric coupling was also expected in Ca$_3$Mn$_2$O$_7$.\cite{Benedek:Prl}

Soon after the theoretical prediction, Oh {\it et al.}'s experiment confirmed the ferroelectricity in Ca$_3$Ti$_2$O$_7$, although the experimental measured $P\sim10\mu$C/cm$^2$ is somewhat lower than the expected value.\cite{Oh:Nm} Similar ferroelectricity was also predicted and later observed in Sr$_3$Sn$_2$O$_7$.\cite{Mulder:Afm,Wang:am17} However, direct experimental measurement of ferroelectricity remains absent for Ca$_3$Mn$_2$O$_7$. In fact, an experimental study on Mn substituted Ca$_3$Ti$_2$O$_7$, i.e. Ca$_3$Ti$_{2-x}$Mn$_x$O$_7$, found suppressed ferroelectricity, i.e. reduced $P$ and lowered transition temperature ($T_{\rm C}$), upon increasing concentration of Mn.\cite{Liu:Apl} The temperature ($T$)-dependent structural evolution of Ca$_3$Mn$_2$O$_7$ is more complex than that of Ca$_3$Ti$_2$O$_7$. For example, a phase coexistence over large $T$ range and the ``symmetry trapping" of a soft mode were observed.\cite{Lobanov:Jpcm,Senn:Prl}

In this work, the high quality polycrystalline samples of Ca$_3$Mn$_2$O$_7$ were prepared by standard solid state reaction method, starting from the highly purified powders of CaCO$_3$ and MnO$_2$. The stoichiometric mixtures were ground and fired at $1200$ $^{\circ}$C for $24$ hours in air. Then the resultant powders were pelletized into a disk of $2$ cm in diameter under $5000$ psi pressure, and sintered at $1350$ $^{\circ}$C for $24$ hours with intermittent grinding step.

The crystal structure was characterized by X-ray diffraction (XRD) with Cu-$K_{\alpha}$ radiation from $207$ K to $673$ K. Using the GSAS Rietveld program,\cite{Larson:laur} the refined crystallographic information of Ca$_3$Mn$_2$O$_7$ are summarized in Table~\ref{Table1}. At high $T$'s (e.g. $673$ K), the structure can be described by the $I4/mmm$ space group. In contrast, the $A2_1am$ space group can well describe the low $T$ (e.g. $207$ K) structure. Such a structural transition from nonpolar $I4/mmm$ to polar $A2_1am$ agrees with previous reports.\cite{Lobanov:Jpcm,Senn:Prl} In the middle $T$ range (e.g. $\sim300-600$ K), the mixture of multiple structures including $I4/mmm$, $A2_1am$, as well as the intermediate $Amam$, are evidenced, further confirming the complicated structural evolution from paraelectric to ferroelectric states.

\begin{figure}
\centering
\includegraphics[width=0.48\textwidth]{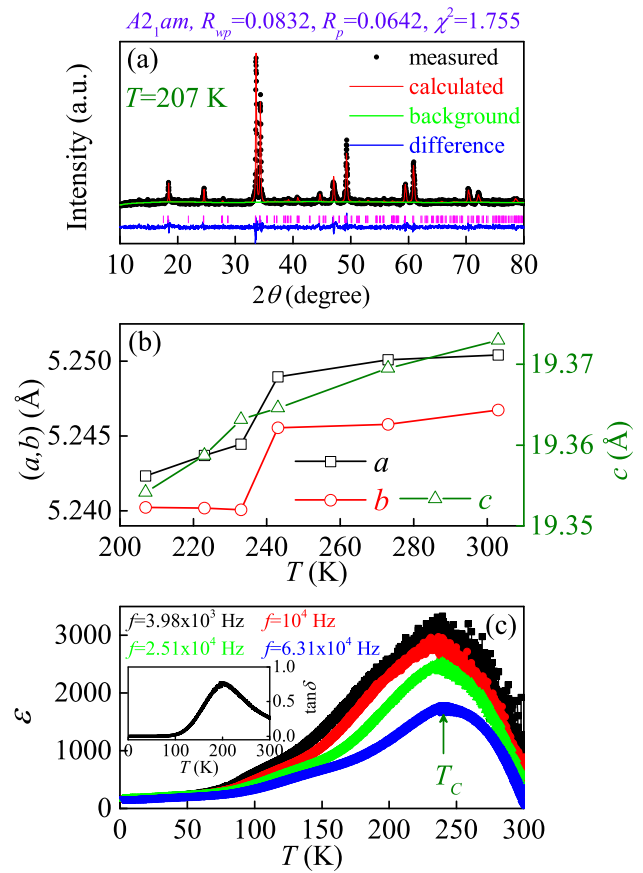}
\caption{(a) Rietveld fitting of the low-$T$ powder XRD spectrum for Ca$_3$Mn$_2$O$_7$. (b) $T$-dependent evolution of lattice constants. (c) $T$-dependent dielectric constant. $f$: measuring frequency. Insert: the dielectric loss.}
\label{Fig1}
\end{figure}

In the following, the low $T$ polar $A2_1am$ structure is studied to verify its multiferroicity. The XRD pattern at $207$ K is shown in Fig.~\ref{Fig1}(a). The refined lattice constants (using the $A2_1am$ space group) as a function of $T$ below $300$ K are shown in Fig.~\ref{Fig1}(b). For lattice constants $a$ and $b$, simultaneous sudden drops are evidenced around $240$ K.

\begin{table}
\centering
\caption{Lattice constants (in unit of \AA) of Ca$_3$Mn$_2$O$_7$ determined by Rietveld analysis at low and high $T$'s (in unit of K).}
\begin{tabular*}{0.48\textwidth}{@{\extracolsep{\fill}}llllllll}
\hline
\hline
$T$  & Group & $a$ & $b$ & $c$ & $R_{\rm WP}$ & $R_{\rm P}$ & ${\chi}^2$\\
$207$ & $A2_1am$ & $5.2423$ &  $5.2402$  & $19.3542$ & $8.32\%$ & $6.42\%$ & $1.755$\\
$673$ & $I4/mmm$ & $3.7096$ & $3.7096$ & $19.5050$ & $8.98\%$ & $6.95\%$ & $1.969$\\
\hline
\hline
\end{tabular*}
\label{Table1}
\end{table}

To measure the dielectric and ferroelectric properties, the sample was polished into thin plate with typical thickness of $0.20$ mm and area of $\sim7.06$ mm$^2$. The gold electrodes were deposited on the top/bottom surfaces. The dielectric constant $\varepsilon$ under different frequencies are measured as a function of $T$ using HP$4294$A impedance analyzer, as shown in Fig.~\ref{Fig1}(c). The variable-$T$ environment is provided by Physical Property Measurement System (PPMS) of Quantum Design, which can cover $1.8-300$ K. A broad peak of $\varepsilon$ is observed at $\sim240$ K, coinciding with the sudden drops of in-plane lattice constants (Fig.~\ref{Fig1}(b)). The peak slightly shifts to the high $T$ direction with increasing frequency. These characteristics seem to suggest relaxor behavior, which is reasonable considering the large $T$ range of phase coexistence.\cite{Lobanov:Jpcm,Senn:Prl} The dielectric loss is very small at low $T$, but becomes considerable large when $T>100$ K due to serious leakage. This may be the reason why previous experiments were failed to directly measure the ferroelectric $P$ at high $T$. Such a large dielectric loss at $\sim240$ K also bring somewhat uncertainness regarding $T_{\rm C}$, which needs further experimental verifications.

\begin{figure}
\centering
\includegraphics[width=0.45\textwidth]{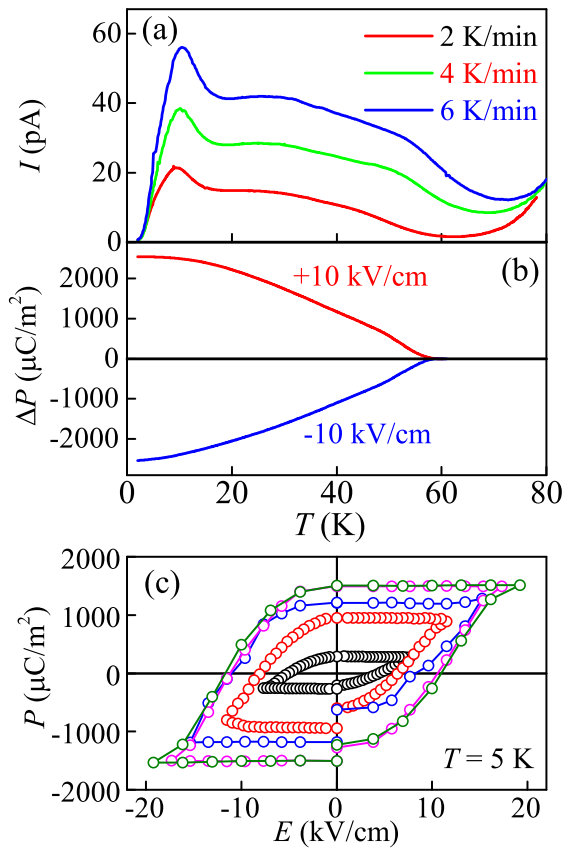}
\caption{(a) Raw data of pyroelectric current $I_{\rm pyro}$ as a function of $T$ at three warming rates $2$, $4$, and $6$ K/min, respectively. (b) Integrated pyroelectric $\Delta P$ (from $2$ K to $60$ K) under positive and negative poling electric fields ($\pm10$ kV/cm). (c) The ferroelectric hysteresis loops measured at $5$ K by the PUND method with different peak voltages.}
\label{Fig2}
\end{figure}

The $T$-dependent ferroelectric $P$ was measured using the pyroelectric current method. In detail, the sample was first poled under a poling electric field from $300$ K to $2$ K. Then the electric field was set to zero, and the sample was electrically short-circuited for several hours at $2$ K in order to exclude possible extrinsic contributions (e.g. trapped charge during the poling process). Then the pyroelectric current ($I_{\rm pyro}$) was collected by heating the sample at different rates of $2$, $4$ and $6$ K/min, as shown in Fig.~\ref{Fig2}(a). All peaks of $I_{\rm pyro}$-$T$ curves are exactly at the same position without any shift. Since the current signal increases rapidly when $T$ is above $60$ K which may be contributed by the extrinsic thermal excitation, the reliable pyroelectric signal in our work is limited to the low $T$ ($<60$ K) region. Noting it does not mean that the ferroelectric $T_{\rm C}$ is $60$ K.

Under the poling field $10$ kV/cm, the integrated pyroelectric $\Delta P$ is about $2500$ $\mu$C/m$^2$ from $60$ K to $2$ K for the polycrystalline sample. And the integrated pyroelectric $\Delta P$ is also independent on the warming rates. The pyroelectric curves are also measured under the positive and negative poling electric fields. The antisymmetrical pyroelectric curves upon the positive/negative poling fields indicate the reversibility of $\Delta P$, as shown in Fig.~\ref{Fig2}(b). All these characteristics imply that the measured signals indeed come from the intrinsic ferroelectricity, although some extrinsic factors may also co-exist and contribute to a portion of $\Delta P$. It should be noted that such $\Delta P$ is only a part of total $P$ since the ferroelectric $T_{\rm C}$ is much higher than $60$ K.

In addition to the pyroelectric measurement, the ferroelectric hysteresis loops have also been measured using the improved Positive-Up-Negative-Down (PUND) method, which can deduct the extrinsic contribution from leakage and capacitance to some extent. The PUND loop at $5$ K is shown in Fig.~\ref{Fig2}(c), which unambiguously demonstrates the ferroelectricity of Ca$_3$Mn$_2$O$_7$. With increasing $T$, the PUND loops gradually shrink and becomes unmeasurable due to serious leakage when $T>28$ K (Fig. S2 of supplementary material).

\begin{figure}
\centering
\includegraphics[width=0.45\textwidth]{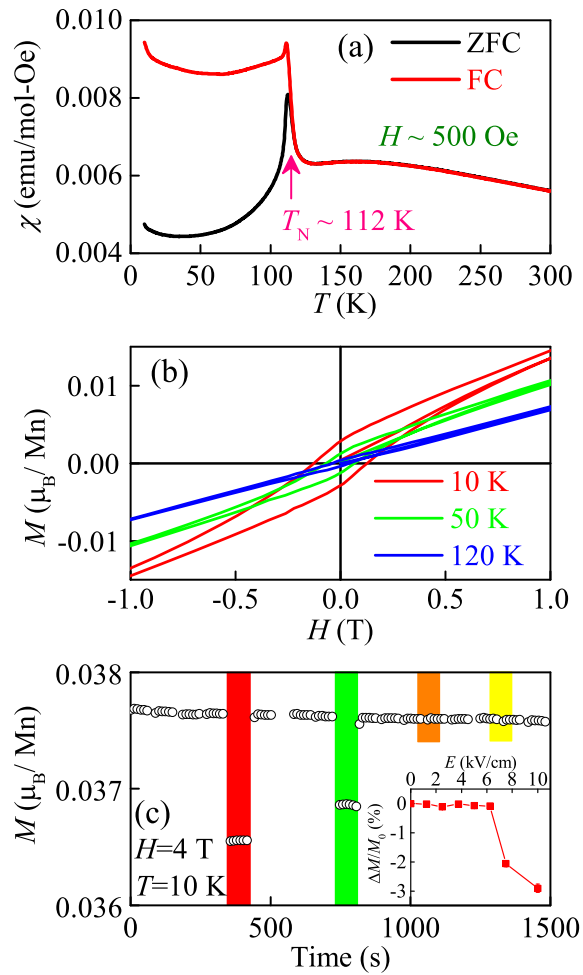}
\caption{(a) The magnetic susceptibilities indicate an antiferromagnetic transition at $112$ K. (b) The magnetic hysteresis loops at different $T$'s. (c) Evolution of $M$ under $4$ T field at $10$ K as a function of time. The color regions denote the periods with electric fields $E$ (red: $10$ kV/cm; green: $7.5$ kV/cm; orange: $5.0$ kV/cm; yellow: $2.5$ kV/cm). Insert: Evolution of $M$ as a function of $E$. The critical field is $\sim6.25$ kV/cm, only beyond which $M$ can be affected.}
\label{Fig3}
\end{figure}

The magnetic susceptibilities ($\chi$'s) as a function of $T$ under the zero-field cooling (ZFC) and field-cooling (FC) modes were measured under a small magnetic field $500$ Oe by the superconducting quantum interference device magnetometer (SQUID) (Quantum Design, Inc.), as shown in Fig.~\ref{Fig3}(a). The peaks of $\chi$'s appear at $112$ K, and the ZFC and FC curves diverge at this point, indicating an antiferromagnetic transition $T_{\rm N}$, in agreement with previous literature.\cite{Lobanov:Jpcm,Zhu:apl12} The specific heat measurement also exhibits a weak anomaly at this $T_{\rm N}$ (Fig. S3 of supplementary material). According to the Curie-Weiss fitting above $T_{\rm N}$, the effective magnetic moment is $3.17$ $\mu_{\rm B}$/Mn, close to the expected value of spin-only magnetic moment ($3$ $\mu_{\rm B}$/Mn) for the high-spin Mn$^{4+}$ ($S_z=\frac{3}{2}$).

Figure \ref{Fig3}(b) shows the magnetic hysteresis loops measured at different $T$'s. Below $T_{\rm N}$, weak FM (wFM) type loops are observed, while a paramagnetic loop is evidenced above $T_{\rm N}$. According to the theoretical prediction, a net magnetization $M\sim0.045$ $\mu_{\rm B}$/Mn can be generated due to the spin canting of antiferromagnetic background.\cite{Benedek:Prl} Here the residual $M$ is about $0.0025$ $\mu_{\rm B}$/Mn at $10$ K. The value of $M$ increases continuously with magnetic field, which is not saturated even under a high field up to $6.5$ T (Fig. S3 of supplementary material).

\begin{figure}
\centering
\includegraphics[width=0.45\textwidth]{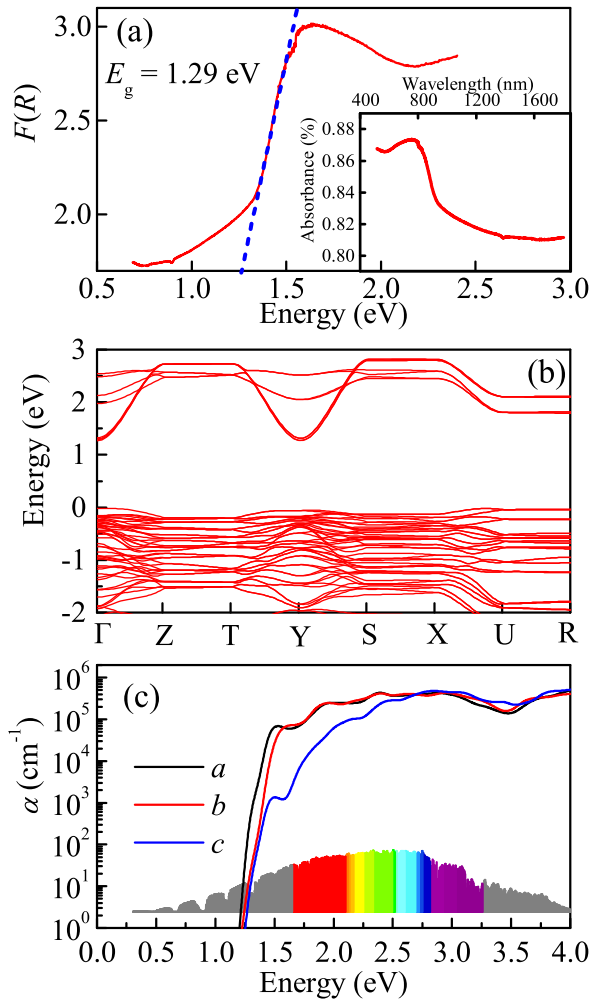}
\caption{(a) The Kubelka-Munk functions $F$($R$) as a function of photon energy. The derived optical band gap is $1.29$ eV. Insert: the absorbance of Ca$_3$Mn$_2$O$_7$. (b) The GGA+$U$ calculated band structure for the ground magnetic state. (c) The calculated light absorption spectra. Different colors are for the electric field components of light along different axis. The standard AM1.5G solar spectrum is shown
on bottom as a reference.\cite{Am1.5}}
\label{Fig4}
\end{figure}

In the original theoretical work, the spin-lattice mediated magnetoelectric coupling was predicted, which might lead to electric-control of $M$.\cite{Benedek:Prl} Here we have investigated the magnetic properties of Ca$_3$Mn$_2$O$_7$ under electric field in the DC excitation mode in SQUID. After the FC (with $4$ T magnetic field and $10$ kV/cm electric field) from $150$ K to $10$ K, $M$ (under $4$ T) is monitored upon the application of electric field. As shown in Fig.~\ref{Fig3}(c), $M$ decreases suddenly when the electric field reaches up to a critical value $E_{\rm C}\sim6.25$ kV/cm. Below this critical value, $M$ is quite robust. Although the Joule heat effect may also suppress $M$, it can not explain the sudden change and critical field. Interestingly, the ferroelectric coercive field at $10$ K is $\sim6$ kV/cm (Fig. S2 of supplementary material), very close to this $E_{\rm C}$, implying the switching of ferroelectric domains  may affect the alignment of magnetic moments. If so, Fig.~\ref{Fig3}(c) can reflect the magnetoelectricity. Further deeper investigation of magnetoelectricity needs the single crystalline samples since the magnetoelectric response is orientation-dependent. Upon a $10$ kV/cm electric field, $M$ decreases for $\sim3\%$.

Recently, photovoltaic effects in ferroelectric (or polar) materials have attracted considerable attentions.\cite{Butler:EES,zhang:prm,Nechache:Nph,Huang:Prb} The most attractive advantage of ferroelectric materials is that the internal electric field built by spontaneous $P$ can promote the separation of photo-generated electrons/holes.\cite{Yuan:Nm} However, for most ferroelectric materials, there are also some disadvantages, including: 1) too large band gaps or indirect band gaps, which prevent the efficient absorption of visible light; 2) low mobility of carriers and too large resistivity, which suppress the photo-generated current. These drawbacks obstruct the applications of ferroelectric materials in the photo-electric field.

However, as stated before, Ca$_3$Mn$_2$O$_7$ is not very insulating, which is disadvantage for ferroelectric measurement but may be advantage for photovoltaic current or other photo applications. To characterize the optical properties of Ca$_3$Mn$_2$O$_7$, the UV-vis-NIR diffuse reflectance spectrum was measured at room $T$ with a Cary $4000-5000-6000i$ UV-vis-NIR spectrophotometer in the $200-2500$ nm wavelength range. The reflectance spectrum was further converted to absorbance with the Kubelka-Munk function.\cite{Kubelka:phy} As shown in Fig.~\ref{Fig4}(a), the Kubelka-Munk function $F$($R$) shows the absorption edge $\sim1.29$ eV (in the infrared light region). The optical band gap for Ca$_3$Mn$_2$O$_7$ has not been reported before. The sample is black with very excellent absorbance of visible light.

To further understand the optical properties, a calculation based on density function theory (DFT) is performed. Although an early DFT calculation had been done for Ca$_3$Mn$_2$O$_7$, they used the pure generalized gradient approximation (GGA) without $U$, which leaded to an unrealistic too small band gap ($\sim 0.3$ eV).\cite{Matar:Prb} Here our DFT calculations are performed using the projector augmented wave (PAW) pseudopotentials as implemented in Vienna {\it ab initio} Simulation Package (VASP) code.\cite{Kresse:Prb,Kresse:Prb96,Kresse:Prb99,Perdew:Prl} To acquire more accurate description of crystalline structure and electron correlation, the revised Perdew-Burke-Ernzerhof for solids (PBEsol) function and the GGA+$U$ method are adopted.\cite{Perdew:Prl,Perdew:Prl08} According to literature,\cite{Benedek:Prl} the on-site Coulomb $U_{\rm eff}=4$ eV is applied to the $3d$ orbital of Mn, using the Dudarev implemention.\cite{Dudarev:Prb} The cutoff of plane wave basis is fixed to $550$ eV. The Monkhorst-Pack $k$-point mesh is $7\times7\times2$.% The standard Berry phase method is adopted to estimate the ferroelectric $P$.\cite{Resta:Rmp}

The DFT band structure shown in Fig.~\ref{Fig4}(b) demonstrates a semiconducting properties with direct band gaps ($\sim1.3$ eV, close to the experimental value) at the $\Gamma$ and $Y$ points. In addition to the appropriate value of band gaps, the conducting bands formed by the $e_{\rm g}$ orbitals are relative wider (comparing with $h$-TbMnO$_3$ \cite{Huang:Prb} and CaOFeS \cite{zhang:prm}), implying a relative large mobility of electrons in the $ab$ plane.

Then, the light absorption spectra are calculated from the imaginary part of the dielectric constant,\cite{Huang:Prb,zhang:prm} as shown in Fig.~\ref{Fig4}(c). The absorption is better for lights with in-plane electric field components. It is quite nature to understand the anisotropic absorption due to the crystal anisotropy. The in-plane hybrid improper ferroelectricity also leads to tiny anisotropy between the $a$ and $b$ axes for the in-plane electric field components. The first peak of absorption appears near $1.3$ eV. Strong absorption in the whole visible light range is found in our calculation. All these characteristics agree with aforementioned experimental results. Further photoelectric measurements on Ca$_3$Mn$_2$O$_7$ are encouraged to check its potential applications as a ferroelectric photovoltaic material.

In summary, the physical properties of Ca$_3$Mn$_2$O$_7$ have been investigated, including its structural property, magnetism, ferroelectricity, magnetoelectricity, and optical property. Measurements of dielectric constants and structure suggest possible ferroelectric $T_{\rm C}$ around $240$ K, although the relaxor like behaviors are found in a quite wide temperature range. The ferroelectric polarization is measured at low temperature by the pyroelectric method as well as the PUND method, the later of which can give the hysteresis loops of electric-polarization. Furthermore, the net magnetization of weak ferromagnetism is evidenced, which can be modulated by electric field beyond the ferroelectric coercive field. Finally, the strong light absorption has been confirmed in both experiment and DFT calculation, implying potential ferroelectric photo-electric applications.

\section{Supplementary Material}
See supplementary material for the structures and more experimental data of Ca$_3$Mn$_2$O$_7$.

This work was supported by the National Key Research Projects of China (Grant No. 2016YFA0300101), the National Natural Science Foundation of China (Grant Nos. 11704109, 11674055, 51431006, 51332006, 51721001, 11504048, 51790491), Research Project of Hubei Provincial Department of Education (Grant No. Q20172501). Most calculations were supported by National Supercomputer Center in Guangzhou (Tianhe II).

\bibliographystyle{apsrev4-1}
\bibliography{../ref3}
\end{document}